\documentclass[aps,pra,twocolumn,showpacs,superscriptaddress]{revtex4-1}

\usepackage{amsfonts,mathrsfs,amsmath,amsthm,amssymb}
\usepackage{bm,times}
\usepackage{graphicx,epsfig}
\usepackage[colorlinks=true,breaklinks=true,linkcolor=blue,citecolor=blue,urlcolor=blue]{hyperref}

\def \tr{ {\rm{Tr}}}
\def \non {\nonumber}

\def \ket {\rangle}
\def \bra {\langle}

\newcommand{\be}{\begin{eqnarray}}
\newcommand{\ee}{\end{eqnarray}}

\makeatletter
    
    \newcommand{\Rmnum}[1]{\expandafter\@slowromancap\romannumeral #1@}
\makeatother

\begin{document}

\title{Certifying quantumness with the classical fidelity threshold}
\author{Long Huang}
\affiliation{College of Physical Science and Technology, Sichuan University, Chengdu 610064, China}
\author{Xiaohua Wu}
\email{wxhscu@scu.edu.cn}
\affiliation{College of Physical Science and Technology, Sichuan University, Chengdu 610064, China}
\author{Tao Zhou}
\email{taozhou@swjtu.edu.cn}
\affiliation{Department of Applied Physics, School of Physical Science and Technology, Southwest Jiaotong University, Chengdu 611756, China}

\date{\today}

\begin{abstract}
For a given ensemble of input and target states, the classical fidelity threshold (CFT) is the maximum valve of the averaged fidelity, and it can be achieved with a measure-and-prepare operation. This quantity can be employed to verify whether the  channel is in the quantum domain or not. In a recent work by Chiribella and Xie [Phys. Rev. Lett. {\bf 110}, 213601 (2013)], it was showed that all the information about the input and target states can be equivalently described by an entangled state and an effective entanglement-braking (EB) channel, and the CFTs can be defined with the Choi matrix of the effective EB channel. Following this idea, the protocol proposed by Fuchs and Sasaki [Quantum. Inf. Comput, {\bf 3}, 377 (2003)] are reformulated in terms of the effective EB channel in this paper, and as applications, the deterministic and probabilistic CFTs for qubit states and the coherent states are derived.
\end{abstract}
\maketitle

\section{introduction}
How to establish reliable quantum channels, which can be applied to transmit and store quantum states faithfully, is a central challenge for
the realization of quantum information processing (QIP).  A distinguishing  class of local operations is the so-called \emph{entanglement-breaking} (EB) channel~\cite{horo1,Ruskai}. It is well known that an operation is EB if and only if it can be written as a \emph{measure-and-prepare} (MAP) scheme that assigns output sates based on the classical data obtained by measuring the input states. When a channel is not a MAP scheme, there exists an entangled state, after evolving under a local channel, can still be applied to transmit nonclassical correlation. Reasonably, one may recognize that the channel is in \emph{quantum domain} if it is not a MAP scheme.

In principle, a quantum channel can usually be described by Choi matrix, and be determined by performing \emph{quantum process tomography} (QPT)~\cite{Nielsenbook,ariano1,ariano2}. However, the complete characterization of the Choi matrix is a non-scalable task, and for the N d-level system, there are about $d^{4N}$ elements to be determined. The task of \emph{quantum benchmark} is to certify a device to be in quantum domain. With a defined figure of merit, one should measure this quantity with the experimental device and   calculate the threshold if the channel is supposed to be a MAP scheme.

A number of benchmarks have been developed in recent years. One widely used figure of merit is the averaged (square) Uhlmann fidelity~\cite{Uhlmann}: Alice prepares a state $\vert\Psi\rangle$ and sends it to Bob via the quantum channel. After receiving the state, Bob will measure the fidelity between the output state and a designed target state $\vert\Psi'\rangle$, say $F^2(\vert\Psi'\rangle,\hat{\rho}^{\mathrm{out}})=\langle \Psi'\vert\hat{\rho}^{out}\vert\Psi'\rangle$, with $F(\vert\Psi'\rangle,\hat{\rho}^{\mathrm{\mathrm{out}}})$ to be Uhlmann fidelity. With a given prior probability of the inputs, one may get a averaged fidelity after sufficient runs of experiment. Meanwhile, one has a corresponding theoretical  task  to calculate the \emph{classical fidelity threshold} (CFT) which is defined as the maximum of the fidelity achieved  by a MAP operation. One may declare that the experimental channel is in the quantum domain, as long as the measured quantity exceeds the CFT.

The theoretical studies on CFTs have a long history. It firstly appeared in the finite-dimension system with input and target states described by discrete variables~\cite{Barnett1,Fuchs, Hunter,Namiki1}, or continuous variables (CVs)~\cite{Pop,Massar,Horo,Brub,Cal}. In recent years, great progress has been achieved in the field for CV encodings of light.  CFTs for the coherent states and squeezed states have also been discussed in Refs.~\cite{ Braun,Ham,Adesso1,Chir,Namiki2,Yang}. Benchmarks for the amplification of coherent states are important for assessing the realization of deterministic~\cite{Pooser} or probabilistic amplifiers~\cite{Kocsis,Fer}, and have been theoretically studied in Refs~\cite{Namiki3,Chir1}. Besides the fidelity benchmark, different benchmarks have been developed~\cite{mor,has2,Has3,Kill,Namiki4,Namiki5}.

In the recent work~\cite{Chir1}, Chiribella and Xie showed that all the information about the input and target states can be equivalently described by an entangled state and an effective EB channel, and the CFTs can be defined with Choi matrix of the effective EB channel. From the general theorem of QPT, it is well known that a quantum channel can also be represented by the process matrix which has a one-to-one correspondence to Choi matrix.  Following the idea of Chiribella and Xie, the protocol to calculate CFTs developed by Fuchs and Sasaki~\cite{Fuchs} can be reformulated in terms of the effective EB channel, and the benchmark can be decided by the process matrix of the effective EB channel. Taking qubit states and coherent states as examples, in this paper, it is shown that the reformulated protocol is a convenient tool to obtain CFTs.

This paper is organized as follows. In Sec.~\ref{sec2}, the general theory of QPT, the Fuchs-Sasaki protocol and the concept of
effective EB channel are briefly discussed. With the reformulated protocol defined in Sec.~\ref{sec3}, CFTs for qubit states and coherent states  are calculated in Sec.~\ref{sec4} and Sec.~\ref{sec5}, respectively. Finally, we end our paper with a short discussion in Sec.~\ref{sec6}.

\section{Preliminary}
\label{sec2}

\subsection{Quantum process tomography}
Before giving a brief review of the general theory about QPT, one can first introduce the convenient tool where a bounded operator in a $d$-dimensional Hilbert space $\mathrm{H}_{d}$ is associated with a vector in an extended Hilbert space $\mathrm{H}_{d}^{\otimes 2}$. Let $A$ to be a bounded operator in $\mathrm{H}_d$, with $A_{ij}=\langle i\vert  A\vert j\rangle$ the matrix elements, an isomorphism between  $A$ and a $d^2$-dimensional vector $\vert A\rangle\rangle$ is defined as
\be
\vert A\rangle\rangle =\sqrt{d} A\otimes \mathrm{I}_{d}\vert S_+\rangle=\sum_{i,j=1}^d A_{ij}\vert ij\rangle,
\ee
where $\vert S_+ \rangle =\frac{1}{\sqrt{d}}\sum_{k=1}^{d}\vert kk\rangle$ is the maximally entangled state in $\mathrm{H}_{d}^{\otimes 2}$, and  $\vert ij\rangle=\vert i\rangle\otimes \vert j\rangle$. This isomorphism offers a one-to-one map between an operator and its vector form. Suppose that $A$ , $B$, and $\hat{\rho}$  are three arbitrary bounded operators in $\mathrm{H}_{d}$, and then
\be
\label{trace}
\mathrm{Tr}(A^{\dagger}B)=\langle\langle A\vert B\rangle\rangle,\ \ 
\vert A\hat{\rho} B\rangle\rangle =A\otimes B^{\mathrm{T}}\vert \hat{\rho} \rangle\rangle,
\ee
with $B^{\mathrm{T}}$ the transpose of $B$.

A quantum channel $\varepsilon$ can be described by a set of Kraus operators $\{E_m\}$,
$\varepsilon(\hat{\rho})=\sum_{m} E_m\hat{\rho} E_m^{\dagger}$, and Choi-Jamiolkowski isomorphism is a useful connection between a quantum channel and a bipartite state
\be
\hat{\chi}_{\varepsilon}:&=&d\cdot\varepsilon\otimes \mathrm{I}_d(\vert S_{+}\rangle \langle S_+\vert),\nonumber\\
&=&\sum_{m}\vert E_m\rangle\rangle\langle\langle E_m\vert,
\ee
where $\hat{\chi}_{\varepsilon}$ is the so-called Choi matrix. From the general theory of QPT, a quantum channel can be equivalently represented by a process matrix,
\be
\label{pmatrix}
\hat{\lambda}_\varepsilon:=\sum_m E_m\otimes E_m^*.
\ee

For an arbitrary input state $\hat{\rho}$, a quantum channel $\varepsilon$ will export a corresponding output state $\varepsilon(\hat{\rho})=\sum_{m} E_m\hat{\rho} E_m^{\dagger}$, and according to Eq.~(\ref{trace}) and Eq.~(\ref{pmatrix}), one can have a compact expression between $\vert  \varepsilon(\hat{\rho})\rangle\rangle$ and $\vert \hat{\rho}\rangle\rangle$,
\be
\vert  \varepsilon(\hat{\rho})\rangle\rangle= \hat{\lambda}_\varepsilon\vert \hat{\rho}\rangle\rangle.
\ee

In order to show that there exists a one-to-one correspndence between Choi matrix and the process matrix, one can introduce the following definition. Let $\vert \Omega\rangle$ be a maximally entangled state in $\mathrm{H}^{\otimes 4}$, $\vert \Omega\rangle=\frac{1}{d} \sum_{i,j=1}^d \vert ij ij\rangle$ with $\vert ijkl\rangle=\vert i\rangle\otimes \vert j\rangle\otimes\vert k\rangle\otimes \vert l\rangle$,
a vector $\vert \Gamma)$ in $\mathrm{H}_{d}^{\otimes 4}$ is associated with a bounded operator $\Gamma$ on $H_d^{\otimes 2}$, with its matrix elements $\Gamma_{ij;kl}\equiv \langle ij\vert \Gamma\vert kl\rangle$, and then, via the isomorphism, one can have $\vert \Gamma)\equiv d\cdot\Gamma\otimes \mathrm{I}_{d}^{\otimes 2}\vert\Omega\rangle=\sum_{i,j,k,l=1}^d \Gamma_{ij;kl}\vert ijkl\rangle$.
For three arbitrary bounded matrices $\Gamma$, $\Delta$, and $\Sigma$ in $\mathrm{H}_{d}^{\otimes 2}$, we can have $\mathrm{Tr}(\Gamma^{\dagger}\Delta)=(\Gamma\vert\Delta),\vert \Gamma\Sigma\Delta)=\Gamma\otimes \Delta^{\mathrm{T}} \vert \Sigma)$.

In the enlarged Hilbert space, a special unitary transformation can be introduced $\hat{\beta}=\sum_{i,j,k,l=1}^{d}\vert ijkl\rangle\langle ikjl\vert$, and it is also a Hermitian operator, $\hat{\beta}=\hat{\beta}^{\dagger}=\beta^{-1}$, and has a nice property that $\beta\vert A\otimes B^*\ket\ket=\big\vert \vert A\rangle\rangle\langle\langle B\vert\big)$. Via $\vert\Gamma^{\beta}\ket\ket=\hat{\beta}\vert \Gamma\ket\ket$, $\Gamma$ can be mapped to be a new operator $\Gamma^{\beta}$, and one can obtain $(A\otimes B^*)^{\beta}=\vert A\rangle\rangle\langle\langle B\vert,\ \ 
(\vert A\rangle\rangle\langle\langle B\vert)^{\beta}=A\otimes B^*$. Therefore, the relationship between Choi matrix and the process matrix~\cite{Wu2}
\be
\label{process}
\hat{\lambda}_\varepsilon^{\beta}=\hat{\chi}_\varepsilon,\ \ \hat{\chi}_\varepsilon^{\beta}=\hat{\lambda}_\varepsilon,
\ee
and another simple relationship useful in the following discussions,
\be
\label{relationship}
\mathrm{Tr}[ \Gamma^{\dagger}\Delta]=\mathrm{Tr}[ (\Gamma^{\beta})^{\dagger}\Delta^{\beta}],
\ee
can be obtained.

A bipartite state $\hat{\rho}_{AB}$ shared by Alice and Bob can always be decomposed as
$\hat{\rho}_{AB}=\varepsilon\otimes \mathbb{I}_d(\vert \hat{\tau}^{1/2}\rangle\rangle\langle\langle\hat{\tau}^{1/2}\vert)$~\cite {wu1},
with $\hat{\tau}$ a certain density matrix. According to the definition of  Choi matrix, $\hat{\rho}_{AB}$ can be also expressed as $\hat{\rho}_{AB}=\mathbb{I}_d \otimes (\hat{\tau}^\mathrm{T})^{1/2}\hat{\chi}_{\varepsilon}\mathbb{I}_d\otimes (\hat{\tau}^\mathrm{T})^{1/2}$.
It offers a method, the so-called ancilla-assisted quantum process tomography (AQPT)~\cite{arino}, to determinate the quantum channel experimentally as follows: Prepare an entangled state $\vert \tau^{1/2}\rangle\rangle$, then after the evolution under $\varepsilon\otimes \mathbb{I}_d$,  measure the output state $\hat{\rho}_{\mathrm{AB}}$ by quantum state tomography, and finally one can have Choi matrix
\be
\label{choi}
\hat{\chi}_{\varepsilon}=\mathbb{I}_d \otimes (\hat{\tau}^\mathrm{T})^{-1/2}\hat{\rho}_{AB} \mathbb{I}_d\otimes (\hat{\tau}^\mathrm{T})^{-1/2}.
\ee

Among  all the quantum channels, we denote a special class referred as the measure-and-prepare  (MAP) channel by $\varepsilon_{\mathrm{MAP}}$. With a set of positive-operator-valued-measure (POVM) operators $\{\hat{\Pi}_y|\ \sum_y \hat{\Pi}_y=\mathbb{I}_d\}$, and a set of pure normalized states $\{\hat{\xi}_y=\vert \xi_y\rangle \langle \xi_y\vert\}$, Choi matrix and the process matrix of a MAP channel $\varepsilon_{\mathrm{MAP}}$ can be expressed as
\be
\hat{\chi}_{\varepsilon_{\mathrm{MAP}}}=\sum_y\hat{\xi}_y\otimes \hat{\Pi}_y^{*},\ \ \hat{\lambda}_{\varepsilon_{\mathrm{MAP}}}=\sum_y\vert \hat{\xi}_y\rangle\rangle\langle\langle\hat{\Pi}_y\vert,
\ee
where $\hat{A}^{*}=(\hat{A}^{\dagger})^\mathrm{T}$ is the complex conjugation of the operator $\hat{A}$.

Now, with $\hat{\chi}_{\varepsilon_{\mathrm{MAP}}}$ defined above, the state $\hat{\rho}_{\mathrm{AB}}=\varepsilon_{\mathrm{MAP}}\otimes \mathbb{I}_d(\vert \hat{\tau}^{1/2}\rangle\rangle\langle\langle\hat{\tau}^{1/2}\vert)$ can be also expressed as $\hat{\rho}_{AB}=\sum_y\hat{\xi}_y\otimes (\hat{\tau}^{1/2}\hat{\Pi}_y\hat{\tau}^{1/2})^{\mathrm{T}}$. Obviously, it is a product state, and therefore,  all MAP channels are indeed EB, $\varepsilon_{\mathrm{MAP}}\equiv\varepsilon_{\mathrm{\mathrm{EB}}}$.

\subsection{The Fuchs-Sasaki protocol}
To verify that a channel is in quantum domain, Alice can prepare a pure state $\vert\Psi_x\rangle$ as the input for the channel $\varepsilon$, and then Bob measures the overlap between the output $\varepsilon(\hat{\Psi}_x)$  and the designed pure state $\vert\Psi'_x\rangle$ usually referred as the target state. After sufficient runs of experiment, a averaged (square Uhlmann) fidelity can be obtained,
\be
F[\varepsilon]=\sum_x p_x\mathrm{Tr}[\hat{\Psi}'_x \varepsilon(\hat{\Psi}_x)],
\ee
with $p_x$ the prior probability for the state $\hat{\Psi}_x=\vert\Psi_x\ket\bra\Psi_x|$. With $\varepsilon_{\mathrm{MAP}}(\hat{\Psi}_x)=\sum_y\mathrm{Tr}[\hat{\Pi}_y\hat{\Psi}_x]\hat{\xi}_y$, the fidelity of a given MAP channel should be
\be
F[\varepsilon_{\mathrm{MAP}}]=\sum_x \sum_y p_x\mathrm{Tr}[\hat{\Pi}_y\hat{\Psi}_x]\mathrm{Tr}[\hat{\xi}_y\hat{\Psi}_x'].
\ee
Then, the CFT, also called quantum benchmark, is defined as the maximum value of $F[\varepsilon_{\mathrm{MAP}}]$,
\be
\label{cft}
F_{\mathrm{c}}=\sup_{\varepsilon_{\mathrm{MAP}}}F[\varepsilon_{\mathrm{MAP}}],
\ee
and \emph{a channel $\varepsilon$ is in quantum domain if $F[\varepsilon]> F_{\mathrm{c}}$}.

Since the operator $\hat{A}_y\equiv\sum_x p_x\mathrm{Tr}[\hat{\Pi}_y\hat{\Psi}_x]\hat{\Psi}'_x $ positive, to achive the maximum value of $F[\varepsilon_{\mathrm{MAP}}]$, $\hat{\xi}_y$ should be fixed as the eigenvector corresponding to the largest eigenvalue $\lambda_1(\hat{A}_y)$ of $\hat{A}_y$. Therefore, $F[\varepsilon_{\mathrm{MAP}}]=\sum_y\lambda_1(\hat{A}_y)$, and a widely-used formula of CFT has the following form
\be
\label{benchmark}
F_{\mathrm{c}}=\sup_{\hat{\Pi}_y}\sum_{y}\lambda_1(\hat{A}_y).
\ee

Next, we shall focus on the protocol developed by Fuchs and Sasaki~\cite{Fuchs}: (a) For a set of input states, define the density matrix $\hat{\tau}$
\be
\hat{\tau}=\sum_x p_x \hat{\Psi}_x;
\ee
(b) With a POVM $\{\hat{\Pi}_y\}$, another set of probability distribution $\{p_y\}$ can be obtained,
\be
p_y=\mathrm{Tr}[\hat{\Pi}_y\hat{\tau}];
\ee
(c) Define a joint probability distribution $p(x,y)=p_x\mathrm{Tr}[\hat{\Psi}_x\hat{\Pi}_y]$, and based on Bayes' rule, the conditional probability
$ p(x\vert y)$ is
\be
p(x\vert y)=\frac{p_x\mathrm{Tr}[\hat{\Psi}_x\hat{\Pi}_y]}{p_y};
\ee
(d) With the conditional probability, a density matrix $\hat{\rho}_y$ can be introduced
\be
\label{desityy}
\hat{\rho}_y=\sum_x p_xp(x\vert y)\hat{\Psi}'_x,
\ee
and the fidelity of the MP can be rewritten as
\begin{equation}
F(\varepsilon_{\mathrm{MAP}})=\sum_y p_y \lambda_1(
\hat{\rho}_y);
\end{equation}
(e) Finally, the CFT is defined as
\begin{equation}
F_{\mathrm{c}}=\sup_{\hat{\rho}_y}\sum_y p_y \lambda_1(
\hat{\rho}_y).
\end{equation}
The above protocol has been proposed for the derivation of Eq.~(\ref{benchmark}) in the original work~\cite{Fuchs}, while in the present work, we shall reformulate it in terms of the effective EB channel and apply it to calculate CFTs.

\subsection{The effective entanglement-breaking channel}
Recently, it was shown that CFT can be calculated in an enlarged Hilbert space with Choi matrix defined~\cite{Chir1}. To have a better understanding of the result, we introduce the following square-root (Sqrt) transformation, which is usually used in quantum state discrimination (QSD),
\be
\hat{S}_x&=&p_x\hat{\tau}^{-1/2}\hat{\Psi}_x\hat{\tau}^{-1/2},
\ee
with
\be
p_x&=&\tr(\hat{\tau}\hat{S}_x).
\ee
This transformation relates the set of states $\{p_x,\hat{\Psi}_x\}$ to the so-called Sqrt POVM $\{\hat{S}_x\}$. All information about the input and target states can be equivalently described by an entangled state $\vert \hat{\tau}^{1/2}\rangle\rangle$ and an effective EB channel, and  this is one of the main ideas in Ref.~\cite{Chir1}. In the form  of our representation, one may at first define the separable state $\hat{\rho}_{\mathrm{AB}}=\sum_x  p_x\hat{\Psi}'_x\otimes\hat{\Psi}^*_x$, and regard it as the final state after performing AQPT for  the EB channel $\varepsilon_{\mathrm{EB}}$, $\hat{\rho}_{\mathrm{AB}}=\varepsilon_{\mathrm{EB}}\otimes \mathbb{I}_d(\vert \hat{\tau}^{1/2}\rangle\rangle\langle\langle \hat{\tau}^{1/2}\vert)$. With Choi matrix of the EB channel $\varepsilon_{\mathrm{EB}}$
\be
\hat{\chi }_{\varepsilon_{\mathrm{EB}}}=\sum_x \hat{\Psi}'_x\otimes\hat{S}^*_x.
\ee
one can certainly come to
$\hat{\rho}_{\mathrm{AB}}=\mathbb{I}_d\otimes (\hat{\tau}^*)^{\frac{1}{2}}\hat{\chi }_{\varepsilon_{\mathrm{EB}}} \mathbb{I}_d\otimes (\hat{\tau}^*)^{\frac{1}{2}}$. Based on this, the deterministic CFT can be defined as~\cite{Chir1},
\be
\label{deterministic}
F_{\mathrm{c}}^{\mathrm{det}}=\sup_{\varepsilon_{\mathrm{MAP}}}\mathrm{Tr}[\hat{\rho}_{\mathrm{AB}}\hat{\chi}_{\varepsilon_{\mathrm{MAP}}}],
\ee
[It should be emphasized that this definition is just a reformulation of Eq.~(\ref{cft}) in the enlarged Hilbert space.] Meanwhile, the probabilistic CFT is defined as~\cite{Chir1}
\be
F_{\mathrm{c}}^{\mathrm{prob}}=\vert\vert \hat{\chi }_{\varepsilon_{\mathrm{EB}}}\vert\vert_{\times},
\ee
where $\vert\vert \hat{B}\vert\vert_{\times}$ denotes the injective cross norm, $\vert\vert \hat{B}\vert\vert_{\times}=\sup_{\vert\vert{\hat\psi}\vert\vert=\vert\vert{\hat\phi}\vert\vert=1}\tr(\hat{\psi}\otimes \hat{\phi} \hat{B})$. Obviously, the probabilistic CFT is the upper-bound of the deterministic one, $F_{\mathrm{c}}^{\mathrm{det}}\leq F_{\mathrm{c}}^{\mathrm{prob}}$.

\begin{figure}
\centering
\includegraphics[scale=0.37]{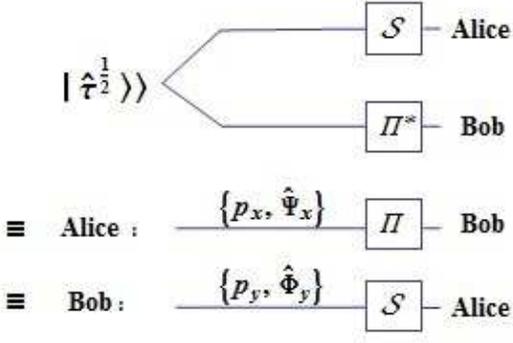}
\caption{\label{fig:epsart} The Bayes' rule can be interpreted as: (1) Alice and Bob simultaneously performs measurement for the entangled state $\vert  \hat{\tau}^{1/2}\rangle\rangle$ with $\{\hat{S}_x\}$ and $\{\hat{\Pi}^*_y\}$, respectively; ({2}) Alice uses the entangled  state as resource to prepare a set of  state $\{p_x, \hat{\Psi}_x\}$ and sends it to Bob, then Bob  performs the measurement $\{\hat{\Pi}_y\}$ on these states;
 and (3) Bob uses the entangled  state
as resource to prepare a set of  state $\{p_y, \hat{\Phi}_y\}$. After receiving the  states,  Alice   will perform the measurement $\{\hat{S}_x\}$.
All of these three types of experiments are equivalent for deciding the joint probability.}
\label{fig1}
\end{figure}

\section{reformulation of the Fuchs-Sasaki protocol}
\label{sec3}
First, an inverse Sqrt transformation can be introduced
\be
\hat{\Phi}_y&=&p_y^{-1}\hat{\tau}^{1/2}\hat{\Pi}_y\hat{\tau}^{1/2},
\ee
and 
\be
\sum_y p_y\hat{\Phi}_y&=&\hat{\tau},
\ee
and this transformation relates the rank-one POVM $\{\hat{\Pi} _y\}$, to the set of pure states $\{p_y, \hat{\Phi}_y\}$. As an application of Eq.~(\ref{relationship}), we have
\be
\label{joint}
\langle\langle \hat{\tau}^{1/2}\vert \hat{S}_x\otimes  \hat{\Pi}_y^*\vert\hat{\tau}^{1/2}\rangle\rangle=p_x\langle\langle \hat{\Pi}_y\vert\hat{\Psi}_x\rangle\rangle=p_y\langle\langle \hat{S}_x\vert \hat{\Phi}_y\rangle\rangle.\non\\
\ee
These relations can be interpreted by Bayes's rule by setting $p(x,y)=\langle\langle \hat{\tau}^{1/2}\vert \hat{S}_x\otimes  \hat{\Pi}_y^*\vert\hat{\tau}^{1/2}\rangle\rangle,\ \ p(y\vert x)=\langle\langle \hat{\Pi}_y\vert\hat{\Psi}_x\rangle\rangle$, and $p(x\vert y)=\langle\langle \hat{S}_x\vert \hat{\Phi}_y\rangle\rangle$, and the physical interpretation is shown in FIG.~\ref{fig1}: (1) Alice and Bob simultaneously perform the measurements for the entangled state $\vert  \hat{\tau}^{1/2}\rangle\rangle$ with $\{\hat{S}_x\}$ and $\{\hat{\Pi}^*_y\}$, respectively; (2) Alice uses the entangled state as resource to prepare a set of  state $\{p_x, \hat{\Psi}_x\}$ and sends it to Bob, and then, Bob performs the measurement $\{\hat{\Pi}_y\}$ on these states; (3) Bob uses the entangled state as resource to prepare a set of state $\{p_y, \hat{\Phi}_y\}$, and after receiving the  states,  Alice will perform the measurement $\{\hat{S}_x\}$. From Eq.~(\ref{joint}) and FIG.~\ref{fig1}, the three types of experiments are equivalent to realize the joint probability.

By jointing above results with the effective EB channel together, Fuchs and Sasaki's protocol can be interpreted like this:
Using the entangled  state $\vert  \hat{\tau}^{1/2}\rangle\rangle$
as resource, Bob  prepares a set of  state $\{p_y, \hat{\Phi}_y\}$  and sends it to Alice via the EB breaking channel $\varepsilon_{\mathrm{EB}}$.
Now, the density matrix $\hat{\rho}_y$, which has been defined in Eq.~(\ref{desityy}), will have a compact form,
 \be
 \hat{\rho}_y=\varepsilon_{\mathrm{EB}}(\hat{\Phi}_y)
\ee
As in Ref.~\cite{Chir1}, $\hat{\rho}_y$ can be calculated in the enlarged Hilbert space
\be
\vert \hat{\rho}_y\rangle\rangle&=& \hat{\lambda}_{\varepsilon_{\mathrm{EB}}}\vert \hat{\Phi}_y\rangle\rangle\\
\hat{\lambda}_{\varepsilon_{\mathrm{EB}}}&=&\sum_x\vert\hat{\Psi}'_x\rangle\rangle\langle\langle\hat{S}_x\vert,
\ee
where $\hat{\lambda}_{\varepsilon_{\mathrm{EB}}}$ is the process matrix of the effective EB channel. Formally, the deterministic CFT in  Eq.~(\ref{deterministic}) can be rewritten as
\be
\label{detercft}
F^{\mathrm{det}}_{\mathrm{c}}=\sup_{\sum_y p_y\hat{\Phi}_y=\hat{\tau}}\sum_{y}p_y\vert\vert\varepsilon_{\mathrm{EB}}(\hat{\Phi}_y)\vert\vert_{\infty},
\ee
with $\vert\vert \hat{A}\vert\vert_{\infty}=\sup_{\vert\vert\hat{\psi}\vert\vert=1}\langle\psi\vert \hat{A}\vert\psi\rangle$ the operator norm, and certainly, $\vert\vert\varepsilon_{\mathrm{EB}}(\hat{\Phi}_y)\vert\vert_{\infty}=\lambda_1(\varepsilon_{\mathrm{EB}}(\hat{\Phi}_y))$. Using Eq~(\ref{relationship}), the probabilistic CFT can also be expressed as
\be
\label{probcft}
F^{\mathrm{prob}}_{\mathrm{c}}=\sup_{\vert\vert\hat{\Phi}\vert\vert=1}\vert\vert\varepsilon_{\mathrm{EB}}(\hat{\Phi})\vert\vert_{\infty}.
\ee

Denote a unitary channel by  $\mathcal{U}$, $\mathcal{U}(\hat{\rho})=U \hat{\rho} U^{\dagger}$, with $U^{\dagger}=U^{-1}$, there should be
$\mathcal{U}\circ \mathcal{U^{\dagger}}=\mathbb{I}_d$. Define a channel $\tilde{\varepsilon}=\mathcal{V}\circ\varepsilon\circ \mathcal{U}^{\dagger}$, and the identity $\varepsilon=\mathcal{V}^{\dagger}\circ\tilde{\varepsilon } \circ\mathcal{U}$ if $\tilde{\varepsilon}$ can be expressed in terms of the process matrices,
\be
\hat{\lambda}_{\tilde{\varepsilon}}=V\otimes V^{*}\hat{\lambda}_{\varepsilon} (U\otimes U^*)^{\dagger}.
\ee
Furthermore, if $\hat{\lambda}_{\tilde{\varepsilon}}=\hat{\lambda}_{\varepsilon}$, we say that the channel $\varepsilon$ is  invariant  under the unitary decomposition with $U$ and $V$.  Assuming that  $\hat{{\Phi}}$ is an input for such an invariant channel, another input $\tilde{{\Phi}}=U\hat{\Phi}U^{\dagger}$ will have a corresponding output $\varepsilon(\tilde{{\Phi}})=V^{\dagger}\varepsilon(\hat{\Phi})V$, and one can have a useful relationship
\be
\label{relation}
\vert\vert\varepsilon(\tilde{{\Phi}}) \vert\vert_{\infty}=\vert\vert\varepsilon(\hat{\Phi}) \vert\vert_{\infty}.
\ee
Based on the results above, one can come to such an ansatz: \emph{If $ \hat{\Phi}$ is the input of an invariant channel, then  $\tilde{{\Phi}}$ should be also included in the set of inputs.}

For simplicity, the target states $\{\tilde{\Psi}'_x\}$ are generated by a density matrix $\hat{\tau}'$ via the inverse Sqrt transformation
\be
\hat{\Psi}'_x=p'^{-1}_x\hat{\tau}'^{1/2}\hat{S}_x \hat{\tau}'^{1/2}, p_x'=\mathrm{Tr}[\hat{S}_x\hat{\tau}'],
\ee
and with the fact that the CFT never changes under the transformations $\hat{\Psi}_x'\rightarrow V\hat{\Psi}_x' V^{\dagger}$, $\hat{\tau}' \rightarrow V\hat{\tau}' V^{\dagger}$, we always fix $\hat{\tau}'$ to be diagonal, $\hat{\tau}'=\sum_{i=m}^{d}\lambda_m \vert m\rangle\langle m\vert$, with $\sum_m\vert m\rangle\langle m\vert=\mathbb{I}_d$.

Before one can carry on, an algebra inequality which is useful in the following discussions will be introduced in the end of this section. In real parameters domain, for $a>0,q_i>0$ and $-a<x_i<a$,  it can be directly verified that
\be
q_1\sqrt{a^2-x_1^2}+q_2\sqrt{a^2-x_2^2}\leq \sqrt{(q_1+q_2)^2a^2-\bar{x}^2},\non
\ee
with the averaged value $\bar{x}=\sum_{i=1}^{2}q_ix_i$. By repeatedly using this inequality, one can obtain
\be
\label{inequality}
\sum_{i=1}^N q_i\sqrt{a^2-x_i^2}\leq \sqrt{(\sum_{i=1}^Nq_i)^2a^2-\bar{x}^2}.
\ee

\section{Qubit case}
\label{sec4}

\subsection{The Bloch vector transformation}
A single-qubit state can be expressed in the Bloch representation, such that the state $\hat{\rho}$ can be written as $\hat{\rho}=\frac{1}{2}(\mathrm{I}_2+ \vec{r}\cdot\vec{\sigma})$ with $\vec{r}$ is a three component real vector and $\vec{\sigma}=(\hat{\sigma}_x,\hat{\sigma}_y,\hat{\sigma}_z)$. Meanwhile, it turns out that an arbitrary trace-preserving quantum operation is equivalent to a map such that
\begin{equation}
\vec{r'}\rightarrow\vec{r}=\eta\vec{r}+{\vec{c}},
\end{equation}
with $\eta$ a $3\times 3$ real matrix, ${\vec{c}}$ a constant vector, and  $\varepsilon(\hat{\rho})=\frac{1}{2}(\mathrm{I}_2+ \vec{r'}\cdot\vec{\hat{\sigma}})$. This is an affine map, mapping the Bloch sphere into itself~\cite{Nielsenbook}, and can be explicitly expressed as
\be
\left(
  \begin{array}{c}
    r'_x \\
    r'_y \\
    r'_z \\
  \end{array}
\right)=\left(
          \begin{array}{ccc}
            \eta_{xx} & \eta_{xy}& \eta_{xz} \\
            \eta_{yx} & \eta_{yy} & \eta_{yz} \\
            \eta_{zx} & \eta_{zy} & \eta_{zz} \\
          \end{array} \right)
          \left(
                       \begin{array}{c}
                         r_x \\
                         r_y \\
                         r_z \\
                       \end{array}
                     \right)+\left(
                               \begin{array}{c}
                                 c_x \\
                                 c_y\\
                                 c_z\\
                               \end{array}
                             \right),\nonumber
\ee
with the coefficients defined as
\be
\eta_{ij}=\frac{1}{2}\langle\langle \hat{\sigma}_j\vert \hat{\lambda}_{\varepsilon}\vert \hat{\sigma}_i\rangle\rangle,c_k=\frac{1}{2}\langle\langle \hat{\sigma}_k
\vert\hat{\lambda}_{\varepsilon}\vert \mathbb{I}_2\rangle\rangle.\nonumber
\ee
Meanwhile, the unitary transformation $U=\exp\{-i\frac{\omega}{2}\vec{\sigma}\cdot \vec{n}\}$ corresponds to a rotation matrix $O(\omega,\vec{n})$ in Bloch representation
\be
\eta\rightarrow O(\omega,\vec{n})\eta O^{-1}(\omega,\vec{\mathbf{n}}), {\vec{c}}\rightarrow O(\omega,\vec{\mathbf{n}}){\vec{c}}.\nonumber
\ee

A rotation along the $\vec{\mathbf{z}}$ direction, which is usually used in the present work, can take the form
\be
O(\omega,\vec{\mathbf{z}})=\left(
                             \begin{array}{ccc}
                               \cos\omega & \sin\omega & 0 \\
                               -\sin\omega & \cos\omega & 0 \\
                               0 & 0 & 1 \\
                             \end{array}
                           \right).
\ee

\begin{figure} \centering
\includegraphics[scale=0.37]{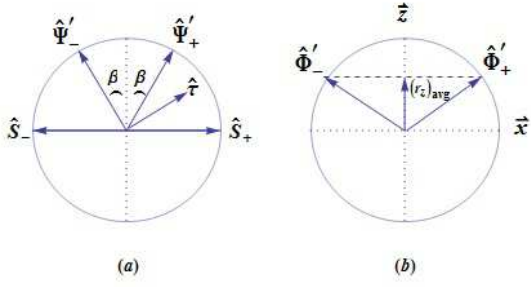}
\caption{\label{fig:epsart} (a) The geometric settings for the Sqrt POVM $\{\hat{\Pi}_{\pm}\}$, the target stats $\hat{\Psi}'_{\pm}$ and the density matrix $\hat{\tau}$.  (b) The optimal inputs for the effective EB channel satisfying the constraint in Eq.~(\ref{constrain}).}
\label{fig2}
\end{figure}

\subsection{A pair of non-orthogonal states}
The case, where the input ensemble consists of a pair of non-orthogonal states while the target state is the same as the input, was first  discussed by Fuchs and Sasaki~\cite{Fuchs}. Later, a more general case, where the target states are different from the inputs, was considered by Namiki \cite{Namiki1}. In the following, it will be shown that the result in Namiki's work can be recovered with the reformulated Fuchs-Sasaki protocol.

As shown in FIG.~\ref{fig2}, one can choose three free parameters, $\alpha$, $\beta$, and $\delta$ satifying
\be
\vert \langle\Psi_+\vert\Psi_{-}\rangle\vert^2=\cos^2\alpha,\vert \langle\Psi'_+\vert\Psi'_{-}\rangle\vert^2=\cos^2\beta, p_{\pm}=\frac{1\pm\delta}{2}.\nonumber
\ee
and the density matrices, Sqrt POVM and the target states can be expressed as
\be
\hat{\tau}&=& \frac{1}{2}(\mathbb{I}_2+\delta\hat{\sigma}_x+\sqrt{1-\delta^2}\cos\alpha\hat{\sigma}_z ),                      \nonumber\\
\hat{S}_{\pm}&=& \frac{1}{2}(\mathbb{I}_2  \pm\hat{\sigma}_x  ),               \nonumber\\
\hat{\Psi}'_{\pm}&=& \frac{1}{2}( \mathbb{I}_2\pm \sin\beta\hat{\sigma}_x+\cos\beta\hat{\sigma}_z).                   \nonumber
\ee
With a simple calculation, we obtain
\be
\eta=\left(
       \begin{array}{ccc}
         \sin\beta & 0 & 0 \\
         0 & 0 & 0 \\
         0 & 0 & 0 \\
       \end{array}
     \right), \vec{\mathbf{c}}=\left(
                                 \begin{array}{c}
                                   0 \\
                                   0 \\
                                   \cos\beta \\
                                 \end{array}
                               \right),
\ee
Taking $\hat{\Phi}=\frac{1}{2}(\mathbb{I}_2+\sin\theta \hat{\sigma}_x+\cos\theta\hat{\sigma}_z)$ as the input of the effective EB channel, the corresponding output should be
\[\varepsilon_{\mathrm{EB}}(\hat{\Phi})=\frac{1}{2}(\mathbb{I}_2+\sin\beta\sin\theta\hat{\sigma}_x+\cos\beta\hat{\sigma}_z),\]
 with its operator norm
\be
\label{norm}
\vert\vert\varepsilon_{\mathrm{EB}}(\hat{\Phi})\vert\vert_{\infty}=\frac{1}{2}(1+\sqrt{1-\sin^2\beta\cos^2\theta}.
\ee
Both the $\eta$ matrix and the shift vector $\vec{\mathbf{c}}$ keep unchanged under the rotation $O(\pi,\mathbf{\vec{z}})$,
 \[\eta=O(\pi,\mathbf{\vec{z}})\eta O^{-1}(\pi,\mathbf{\vec{z}}), \vec{\mathbf{c}}=O(\pi,\mathbf{\vec{z}})\vec{\mathbf{c}}.\]
and based on the results above, we may suppose that the inputs should come in pair:
\[\hat{\Phi}_{\pm}(\theta_i)=\frac{1}{2}(\mathbb{I}_2\pm\sin\theta_i \hat{\sigma}_x+\cos\theta_i\hat{\sigma}_z).\]

 Now, let us return to the CFT defined in Eq.~(\ref{detercft}): For a density matrix $\hat{\tau}$, one may have an arbitrary decomposition of it, $\hat{\tau}=\sum_y p_y\hat{\Phi}_y$, then calculates the average fidelity $F(\varepsilon_{\mathrm{EB}})$, and finally finds out the optimum value of it. For the present case, we use $p_{\pm i}$ to denote the probability for the sates  $\hat{\Phi}_{\pm}(\theta_i)$ and define
 $q_i=p_{+i}+p_{-i}$. Certainly $\sum_{1=1}^{N}q_i=1 $. Due to the constraint $ \hat{\tau}=\sum_{i=1}^N (p_{+i}\hat{\Phi}_{+}(\theta_i)+p_{-i}\hat{\Phi}_{-}(\theta_i))$,
 there should be
\be
\label{constrain}
(r_z)_{\mathrm{avg}}:=\sum_{i=1}^Nq_i\cos\theta_i=\sqrt{1-\delta^2}\cos\alpha.
\ee
With the fact that  $\vert\vert\varepsilon_{\mathrm{EB}}(\hat{\Phi(+})\vert\vert_{\infty}= \vert\vert\varepsilon_{\mathrm{EB}}(\hat{\Phi}_{-})\vert\vert_{\infty}$, for the arbitrary decomposition of $\hat{\tau}$ defined above, we shall get
\be
F(\varepsilon_{\mathrm{EB}})=\frac{1}{2}(1+\sum_{i}q_i\sqrt{1-\sin^2\beta\cos^2\theta_i}).
\ee
Using the inequality in Eq.~(\ref{inequality}), one can obtain
\be
F(\varepsilon_{\mathrm{EB}})\leq\frac{1}{2}(1+\sqrt{1-\sin^2\beta (r_z)^2_{\mathrm{avg}}})\nonumber
\ee
It is easy to verify that the upper bound is tight, and it can be attained when the pair of states
\[\hat{\Phi}_{\pm}=\frac{1}{2}(\mathbb{I}_2\pm\sqrt{1-(r_z)^2_{\mathrm{avg}}} \hat{\sigma}_x+(r_z)_{\mathrm{avg}}\hat{\sigma}_z).\]
is the inputs for the EB channel. Therefore, the CFT is
\be
F^{\mathrm{det}}_\mathrm{c}=\frac{1}{2}(1+\sqrt{1-\sin^2\beta(1-\delta^2)\cos^2\alpha}).
\ee

The present case is special in the sense that when the target states are orthogonal, the CFT equals to the success probability
 $P_{\mathrm{succ}}$ for the discrimination between the inputs with the minimum-error (ME) stragety,
\be
P_{\mathrm{succ}}=\frac{1}{2}(1+\sqrt{1-4p_+p_-\vert\langle\Psi_+\vert\Psi_-\rangle\vert^2},
\ee
the well-known Helstrom bound~\cite{Helstrom}. Moreover, the POVM $\{\hat{\Pi}_{\pm}\}$ , which does not depend on the choice of $\beta$,
is the same as the POVM for the ME discrimination~\cite{Fuchs, Namiki1}.

The probabilistic CFT defibed above is more easily to calculate than the deterministic one, and with the inequality, $F_\mathrm{c}^{\mathrm{prob}}\geq F_\mathrm{c}^{\mathrm{det}}$, the criterion $F[\varepsilon]\geq F_\mathrm{c}^{\mathrm{prob}}$ may be employed to verify that the channel $\varepsilon$ is in quantum domain or not. This is one of the advantages for the probabilistic CFT. However, for some cases, the criterion does not work well since the probabilistic CFT may approach 1. For example, if the input for the effective EB channel is selected from the set $\{\frac{1}{2}(\mathbb{I}_2\pm\hat{\sigma}_x)\}$, by Eq.~(\ref{constrain}), one can have
\be
F^{\mathrm{prob}}_{\mathrm{c}}=1.
\ee

\subsection{Symmetric states}

\begin{figure}
\centering
\includegraphics[scale=0.37]{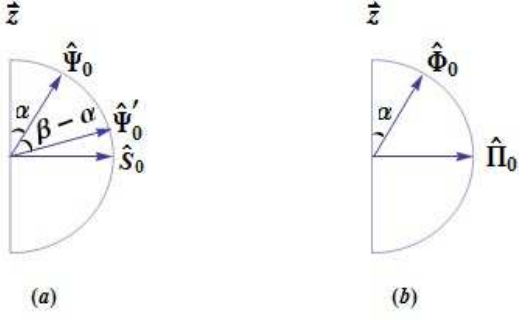}
\caption{\label{fig:epsart} (a) All the states $\hat{\Phi}_x$ ($\hat{\Phi}'_x$) can be generated with the fixed state $\hat{\Psi}_0$ ($\hat{\Psi}'_0$) with the rotation defined in context. In a similar way, the Sqrt POVM $\hat{S}_x$ are also generated from $\hat{S}_0$. (b) The optimal POVM $\{\hat{\Pi}_y\}$  and $\{\Phi_y\}$ can be generated by $\hat{\Pi}_0$ and $\hat{\Phi}_0$, respectively.}
\label{fig3}
\end{figure}

From the definition of the deterministic CFT in Eq.~(\ref{deterministic}), if the bipartite separable states $\hat{\rho}_{\mathrm{AB}}$ are the same,  one can still have the same CFT for the case where the input (or target) states are different. This conclusion has been pointed in previous works, and here an example is give as follows.

Consider that $N$ states $\hat{\Psi}_i=\frac{1}{2}(\mathbb{I}_2+\vec{\hat{\sigma}}\cdot\vec{r}_i)$ are taken as the input with equal probabilities $p_i=1/N$, where each $\vec{r}_i$ are obtained by rotationg a fixed vector $\vec{r}_0$,
\be
\vec{r}_i=O(\omega_i,\vec{\mathbf{z}})\left(
                                  \begin{array}{c}
                                    \sin\alpha \\
                                    0 \\
                                    \cos\alpha \\
                                  \end{array}
                                \right), \omega_i=\frac{i\cdot2\pi}{N}.\nonumber
\ee
In a similar way, the target state can be represented by its Bloch $\vec{r}'_i$
\be
\vec{r}'_i=O(\omega_i,\vec{\mathbf{z}})\left(
                                  \begin{array}{c}
                                    \sin\beta \\
                                    0 \\
                                    \cos\beta \\
                                  \end{array}
                                \right), \omega_i=\frac{i\cdot2\pi}{N}.\nonumber
\ee
The special case with  $\alpha=\beta$ has been discussed in Ref.~\cite{Barnett1}. With the density matrices
\be
\hat{\tau}=\frac{1}{2}(\mathbb{I}_2+\cos\alpha\hat{\sigma}_z), \hat{\tau}'=\frac{1}{2}(\mathbb{I}_2+\cos\beta\hat{\sigma}_z), \nonumber
\ee
the Sqrt POVM $\{\hat{S}_i\}$ can be expressed as
\be
\label{sqrtpovm}
\hat{S}_i=\frac{1}{N}(\mathbb{I}_2+\cos\omega_i\hat{\sigma}_x+\sin\omega_i\hat{\sigma}_y).
\ee
The geometric settings for the input and targets states are depicted in FIG.~\ref{fig3}.

Based on the denotation above, one can obtain
\be
\eta=\left(
       \begin{array}{ccc}
         \frac{1}{2}\sin\alpha & 0 & 0 \\
         0 & \frac{1}{2}\sin\alpha &0 \\
         0 & 0 & 0 \\
       \end{array}
     \right),{\vec{c}}=\left(
                          \begin{array}{c}
                            0 \\
                            0\\
                            \cos\beta\\
                          \end{array}
                        \right).
\ee
which does not depend on the actual number $N$.

Now, for an arbitrary input state
\[\hat{\Phi}(\theta,\phi)=\frac{1}{2}(\mathbb{I}_2+\sin\theta\cos\phi\hat{\sigma}_x+\sin\theta\sin\phi\hat{\sigma}_y+\cos\theta\hat{\sigma}_z),\]
for the effective EB channel, the output should be
\[\varepsilon_{\mathrm{EB}}(\hat{\Phi})=\frac{1}{2}[\mathbb{I}_2+\frac{1}{2}\sin\alpha\sin\theta(\cos\phi\hat{\sigma}_x+\sin\phi\hat{\sigma}_y)+
\cos\beta\hat{\sigma}_z],\]
and the operator norm of the output takes the form
\be
\vert\vert \varepsilon_{\mathrm{EB}}(\hat{\Phi})\vert\vert_{\infty}=\frac{1}{2}(1+\sqrt{\cos^2\beta+\frac{1}{4}\sin^2\alpha(1-\cos^2\theta)}).\nonumber
\ee
For an arbitrary decomposition of the density operator $\hat{\tau}=\sum_m\sum_{n}p_{mn}\hat{\Phi}(\theta_m,\phi_n)$, where $p_{mn}$ are the probabilities for the states $\hat{\Phi}(\theta_m,\phi_n)$, it leads to
\be
(r_z)_{\mathrm{avg}}=:\sum_m q_m \cos\theta_m=\cos\alpha,
\ee
with the denotation $q_m=\sum_{n}p_{mn}$. The average fidelity is obtained as
\be
2F(\varepsilon_\mathrm{{EB}})-1=\sum_mq_m \sqrt{\cos^2\beta+\frac{1}{4}\sin^2\alpha(1-\cos^2\theta_m)}),\nonumber
\ee
and using the inequality in Eq.~(\ref{inequality}), the deterministic CFT become
\be
F^{\mathrm{det}}_{\mathrm{c}}=\frac{1}{2}(1+\sqrt{\cos^2\beta+\frac{1}{4}\sin^4\alpha}).
\ee
It can be achieved by a set of states $\{p_n, \Phi_n\}$,
\[\hat{\Phi}_n=\frac{1}{2}(\mathbb{I}_2+\sin\alpha(\cos\phi_n\hat{\sigma}_x+\sin\phi_n\hat{\sigma}_y)+\cos\alpha \hat{\sigma}_z).\]
with the constraints $\sum_n p_n\cos\phi_n=0, \sum_np_n\sin\phi_n=0$. In terms of the POVM $\{\hat{\Pi}_n\}$,
\[\hat{\Pi}_n=\frac{1}{2}(\mathbb{I}_2+\cos\phi_n\hat{\sigma}_x+\sin\phi_n\hat{\sigma}_y).\]
Obviously, the optimal POVM is not unique. As a special case, the Sqrt POVM $\{\hat{S}_i\}$ in Eq.~(\ref{sqrtpovm}), which is the optimal POVM
for ME discrimination for the inputs $\{\hat{\Psi}_i\}$, belongs to set  $\{\hat{\Pi}_n\}$ above.

Finally, the probabilistic CFT is
\be
F^{\mathrm{prob}}_\mathrm{c}=\frac{1}{2}(1+\sqrt{\cos^2\beta+\frac{1}{4}\sin^2\alpha}),
\ee
which can be achieved if the Bloch vector of input state lies in the $\vec{\mathbf{x}}-\vec{\mathbf{y}}$ plane.

\subsection{Mirror symmetric states}

\begin{figure} \centering
\includegraphics[scale=0.37]{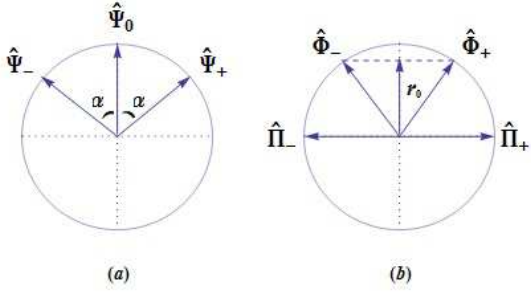}
\caption{\label{fig:epsart} (a) The settings for the input states $\{\hat{\Psi}_x\}$. (b) The optimal POVM $\{\hat{\Pi}_{\pm}\}$ just has two elements although there are three input states. The states $\hat{\Phi}_{\pm}$, the inputs of the effective EB channel, should satisfy the constraints in Eq.~(\ref{cons}).}
\label{fig4}
\end{figure}

As mentioned above, there exist many equivalent ways to define CFT. In Eq.~(\ref{benchmark}), the approach to CFT is realized by finding out the optimal POVM. In the above example, where the probabilities for each input state $\hat{\Psi}_x$ are the same, it has been shown that the Sqrt measurement $\{\hat{S}_x\}$ associated with the input ensemble is optimal. Therefore, one may guess that Sqrt measurement is always optimal when input states are prepared with equal probabilities. Here, we shall provide a counter example.

Shown in FIG.~\ref{fig4}, the input ensemble consists of three mirror symmetric qubit states, and for simplicity, it is assumed that the target states are the same as the inputs, $\hat{\Psi}'_x=\hat{\Psi}_x$ with $x=0,\pm$,
\be
\hat{\Psi}_0&=&\frac{1}{2}(\mathbb{I}_2+\hat{\sigma}_z),\nonumber\\
\hat{\Psi}_{\pm}&=& \frac{1}{2}(\mathbf{I}_2\pm\sin\alpha\hat{\sigma}_x+\cos\alpha\hat{\sigma}_z),\nonumber
\ee
where the prior probability $p_0=p_{\pm}=1/3$.  A more general case, where $p_0\neq p_{\pm}$, has been discussed in Ref.~\cite{Hunter}.
By jointing these states with the density operator,
\be
\label{cons}
\hat{\tau}=\frac{1}{2}(\mathbb{I}_2+r_0\hat{\sigma}_z), r_0=\frac{1}{3}(1+2\cos\alpha),
\ee
one can obtain
\be
\eta=\left(
       \begin{array}{ccc}
         \eta_{xx} & 0 & 0 \\
         0 & 0 & 0 \\
         0 & 0 & \eta_{zz} \\
       \end{array}
     \right), {\vec{c}}=\left(
                         \begin{array}{c}
                           0 \\
                           0 \\
                           c_z \\
                         \end{array}
                       \right),\nonumber
\ee
with the parameters $\eta_{xx}$, $\eta_{zz}$ and $c_z$,
\be
\eta_{xx}&=& \frac{2\sin^2\alpha}{3\sqrt{1-r_0^2}},     \nonumber\\
\eta_{zz}&=& \frac{2r_0\sin^2\alpha}{3\sqrt{1-r_0^2}},       \nonumber\\
c_z&=&  \frac{1+2\cos^2\alpha-3r_0^2}{3(1-r_0^2)}.             \nonumber
\ee
Here, $ 0<\alpha\leq \pi/2$, and based on this, one can come to
\be
\eta_{zz}< \eta_{xx}.
\ee

Now, the input state $\hat{\Phi}(\theta)=\frac{1}{2}(\mathbb{I}_2+\sin\theta\hat{\sigma}_x+\cos\theta\hat{\sigma}_z)$ for the EB channel will produce an output state
\[\varepsilon_{\mathrm{EB}}(\hat{\Phi})=\frac{1}{2}[\mathbb{I}_2+\eta_{xx}\sin\theta\hat{\sigma}_x+(\eta_{zz}\cos\theta+c_z)\hat{\sigma}_z],\]
and we can obtain the operator norm
\be
\vert\vert\varepsilon_{\mathrm{EB}}(\hat{\Phi})\vert\vert_{\infty}=\frac{1}{2}(1+\sqrt{a^2-(b\cos\theta-c)^2},\nonumber
\ee
with $a^2=\eta_{xx}^2+\eta_{zz}^2+c^2$, $b=\sqrt{\eta_{xx}^2-\eta_{zz}^2}$, and $c=\frac{c_{z}\eta_{zz}}{\sqrt{\eta_{xx}^2-\eta_{zz}^2}}.$
Similar to the pair of linearly independent inputs, one can come to
\be
F_\mathrm{c}^{\mathrm{det}}=\frac{1}{2}(1+\sqrt{a^2-(br_0-c)^2},
\ee
which can be attained when the pair of states $\hat{\Phi}_{\pm}=\frac{1}{2}(\mathbb{I}_2\pm\sqrt{1-r_0^2}\hat{\sigma}_x +r_0\hat{\sigma}_z)$ are taken as the inputs for the EB channel. The POVM operators $\{\Pi_{\pm}=\frac{1}{2}(\mathbb{I}_2\pm\hat{\sigma}_x)\}$ are optimal, and for $0<\alpha\leq \pi/4$, this POVM is also optimal for ME discrimination~{\cite{Hunter}.

The probabilistic CFT is the maximum value of $\vert\vert\varepsilon_{\mathrm{EB}}(\hat{\Phi})\vert\vert_{\infty}$
\begin{equation}
F^{\mathrm{prob}}_\mathrm{c}=\frac{1}{2}(1+\vert a\vert),
\end{equation}
which can be easily acquired when $\cos\theta=c/b$.

\subsection{Two pairs of orthogonal states}

In the examples discussed above, the deterministic CFTs are different from the corresponding probabilistic ones. However, in some cases, the two kinds of CFTs may have the same value. In quantum key distribution, two pairs of orthogonal states are usually used to encode information, and the following example is originated from this task.

As shown in FIG.~\ref{fig5}, the ensemble of input states consists of pairs of orthogonal states,
\be
\hat{\Psi}_1&=&\frac{1}{2}(\mathbb{I}_2+\sin\alpha\hat{\sigma}_x+\cos\alpha\hat{\sigma}_z),\nonumber\\
\hat{\Psi}_2&=&\frac{1}{2}(\mathbb{I}_2-\sin\alpha\hat{\sigma}_x-\cos\alpha\hat{\sigma}_z),\nonumber\\
\hat{\Psi}_3&=&\frac{1}{2}(\mathbb{I}_2-\sin\alpha\hat{\sigma}_x+\cos\alpha\hat{\sigma}_z),\nonumber\\
\hat{\Psi}_4&=&\frac{1}{2}(\mathbb{I}_2+\sin\alpha\hat{\sigma}_x-\cos\alpha\hat{\sigma}_z), \nonumber
\ee
with an equal prior probability $p_i=1/4$. The target states are defined by replacing $\alpha$ with $\beta$. For the density matrices $\hat{\tau}=\hat{\tau}'=\frac{1}{2}\mathbb{I}_2$, one can obtain
\be
\eta=\left(
      \begin{array}{ccc}
        \eta_{xx} & 0& 0\\
        0 & 0 & 0 \\
        0 & 0 & \eta_{zz} \\
      \end{array}
    \right), \vec{\mathbf{c}}=0. \nonumber
\ee
with the coefficients $\eta_{xx}=\sin\alpha\sin\beta,\eta_{zz}=\cos\alpha\cos\alpha$.

The output of the input state $\hat{\Phi}(\theta,0)$ for EB channel is
 \[\varepsilon_{\mathrm{EB}}({\hat{\Phi}})=\frac{1}{2}(\mathbb{I}_2+\eta_{xx}\hat{\sigma}_x+\eta_{zz}\hat{\sigma}_z),\]
and the operator norm is
\be
\vert\vert \varepsilon_{\mathrm{EB}}({\hat{\Phi}})\vert\vert_{\infty}=\frac{1}{2}(1+\sqrt{\eta_{xx}^2\sin^2\theta+
\eta_{zz}^2\cos^2\theta}).\nonumber
\ee
\begin{figure} \centering
\includegraphics[scale=0.3]{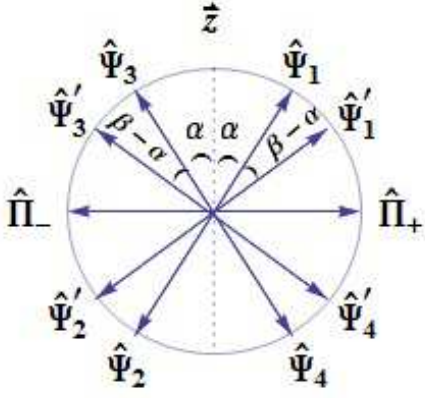}
\caption{\label{fig:epsart} Geometric settings for the input and target states. The Bloch vectors for the optimal POVM $\{\hat{\Pi}_{\pm}\}$ are along the directions $\pm \vec{x}$.}
\label{fig5}
\end{figure}

In practice, one may calculate the probabilistic CFT, and then check whether this CFT, the upper-bound of the deterministic one, is tight or not.  For the present case, the probabilistic   CFT can be easily obtained
\be
F^{\mathrm{prob}}_\mathrm{c}=\{
               \begin{array}{c}
                 \frac{1}{2}(1+\vert\eta_{xx}\vert),   \cos(\alpha-\beta)\cos(\alpha+\beta)< 0 \\
                 \frac{1}{2}(1+\vert\eta_{zz}\vert),  \cos(\alpha-\beta)\cos(\alpha+\beta)> 0  \\
               \end{array},\non\\
\ee
and the above probabilistic CFT is indeed the tight-bound of the deterministic one. For the condition $\cos(\alpha-\beta)\cos(\alpha+\beta)< 0$, \[F^{\mathrm{det}}_\mathrm{c}=\frac{1}{2}(1+\vert\eta_{xx}\vert),\] since that $F^{\mathrm{prob}}_\mathrm{c}$ can be achieved with the optimal POVM $\{\hat{\Pi}_{\pm}=\frac{1}{2}(\mathbb{I}_2+\hat{\sigma}_x)\}$, and for $\cos(\alpha-\beta)\cos(\alpha+\beta)> 0$, the deterministic CFT is
\[F^{\mathrm{det}}_\mathrm{c}=\frac{1}{2}(1+\vert\eta_{zz}\vert),\] with the optimal POVM $\{\hat{\Pi}_{\pm}=\frac{1}{2}(\mathbb{I}_2+\hat{\sigma}_z)\}$. The optimal POVM in the ME discrimination for the input ensemble is $\{\hat{\Pi}_i=\frac{1}{2}\hat{\Psi}_i\}_{i=1}^{4}$, and it is different from the optimal one in the calculation of deterministic CFT.

\section{continuous-variable cases}
\label{sec5}

In this sections, our formulations will be generalized from discrete-variable case to the continuous-variable case by the substitution. First, we will consider the uniform set of input states over a $d$-dimensional Hilbert space and the target state has the same form as its corresponding input.
The bipartite state $\hat{\rho}_{\mathrm{AB}}$ can be generalized as
\be
\hat{\rho}_{\mathrm{AB}}=\int d\mu({\hat{\Psi}})\hat{\Psi}\otimes \hat{\Psi}^*.
\ee
with $d\mu({\hat{\Psi}})$ the Haar measure.  By some simple algebra, in can be known that $\hat{\rho}_{\mathrm{AB}}$ is the separable Werner states~\cite{Werner},
\be
\hat{\rho}_{\mathrm{AB}}=\hat{\rho}_{\mathrm{Werner}}^{\mathrm{sep}}\equiv\frac{1}{d(d+1)}(\mathbb{I}_d\otimes \mathbb{I}_d+ \vert \mathbb{I}_d\rangle\rangle\langle\langle \mathbb{I}_d\vert).
\ee
With $\hat{\tau}=\frac{1}{d}\mathbb{I}_d$, and according to Eq.~(\ref{choi}), the Choi matrix of the effective EB channel can be expressed as
\be
\hat{\chi}_{\varepsilon_{\mathrm{EB}}}=d\cdot\hat{\rho}_{\mathrm{Werner}}^{\mathrm{sep}}.
\ee
Via Eq.~(\ref{process}), the process matrix is
\be
\hat{\lambda}_{\varepsilon_{\mathrm{EB}}}=d\cdot\hat{\rho}_{\mathrm{Werner}}^{\mathrm{sep}},
\ee
For a fixed sate $\vert\Phi_0\rangle$, an arbitrary state $\vert \Phi\rangle$ can be generated through a unitary transformation $U$ on this state, $\vert\Phi\rangle=U\vert\Phi_0\rangle$, and with the invariant property of the effective EB channel, one can have \[ \hat{\lambda}_{\varepsilon_{\mathrm{EB}}}=U\otimes U^*\hat{\lambda}_{\varepsilon_{\mathrm{EB}}}(U\otimes U^*)^{\dagger}\].
Further more, with  Eq.~(\ref{relation}), we have
\be
\vert\vert\varepsilon_{\mathrm{EB}}(\hat{\Phi}) \vert\vert_{\infty}=\vert\vert\varepsilon_{\mathrm{EB}}(\hat{\Phi}_0) \vert\vert_{\infty},\nonumber
\ee
and the probabilistic CFT in Eq.~(\ref{probcft}) is
\be
F^{\mathrm{prob}}_\mathrm{c}=\vert\vert\varepsilon_{\mathrm{EB}}(\hat{\Phi}_0 )\vert\vert_{\infty}.
\ee
For an input state $\hat{\Phi}_0$ for the effective EB channel, the output state is $\varepsilon_{\mathrm{EB}}(\hat{\Phi}_0)=\frac{1}{d+1}(\hat{\Phi}_0+\mathbb{I}_d)$, and then $\vert\vert\varepsilon_{\mathrm{EB}}(\hat{\Phi}_0 )\vert\vert_{\infty}=\frac{2}{d+1}$. Therefore, the probabilistic CFT is
\be
F^{\mathrm{prob}}_\mathrm{c}=\frac{2}{d+1},
\ee
and this is a result in previous works~\cite{Horo,Brub}. This CFT is a tight-bound of the deterministic CFT, and can be attained with any set of rank-one POVM $\{\hat{\Pi}_x\}$.

The classical fidelity threshold for the ensemble of coherent states, which have a Gaussian distribution, was first conjectured by
Braunstein, Fuchs and Kimble~\cite{Braun}. The optimality of this guessed CFT was proven by Hammerer \emph{et al}.~\cite{Ham}. Later, the CFTs for deterministic amplification and and attenuation were put forward by Namik \emph{et al.}~\cite{Namiki3}.  Recently, Chiribella and Xie derived the quantum benchmark for probabilistic amplification of coherent states~\cite{Chir1}. It is interesting that the probabilistic CFT coincides with the deterministic one obtained in Ref.~\cite{Namiki3}, and this result can also be obtained with a self-contained formalism developed by Yang, Chiribella and Adesso~\cite{Yang}. In the present section, we shall show that the probabilistic CFT by Chiribella and Xie, can be also obtained with the reformulated Fuchs-Sasaki protocol. Before one can give such a derivation, we shall introduce some useful results about the coherent states.

First, the displacement operator is defined as
\be
\hat{D}(\alpha)=\exp\{\frac{-\vert\alpha\vert^2}{2}\}\exp\{-\alpha^*\hat{a}^\}\exp\{\alpha\hat{a}^{\dagger}\},
\ee
which is a unitary operator $\hat{D}^{\dagger}(\alpha)=\hat{D}(-\alpha)=[\hat{D}(\alpha)]^{-1}$,
and the displacement operators satisfy a simple multiplication law,
\be
\label{mutiplication}
\hat{D}(\alpha)\hat{D}(\beta)=\hat{D}(\alpha+\beta)\exp\{\frac{1}{2}(\alpha\beta^*-\alpha^*\beta)\}.
\ee
For a complex number $\alpha$, the coherent state $\vert \alpha\rangle$ is defined by
\be
\vert \alpha\rangle&=&\hat{D}(\alpha)\vert 0\rangle,\nonumber\\
&=&\exp\{\frac{-\vert \alpha\vert^2}{2}\}\sum_{n=0}^{\infty}(n!)\alpha^n\vert n\rangle\langle n\vert.
\ee
The thermal state is defined as
\be
\label{thermal}
\hat{T}(0, t)=\frac{1}{1+\langle n\rangle}\sum_{n=0}^{\infty}\bigg(\frac{\langle n\rangle}{1+\langle n\rangle}\bigg)^n \vert n\rangle\langle n\vert,
\ee
where the mean-number of quanta is
\be
\label{mean}
\langle n\rangle=-\frac{1}{2}(1+t),
\ee
and then the density operator $\hat{T}(\gamma, t)$ is defined as
\be
\hat{T}(\gamma, t)=\hat{D}(\gamma)\hat{T}(0, t)\hat{D}^{\dagger}(\gamma), \ \hat{T}(\beta,-1)=\vert\beta\rangle\langle \beta\vert.
\ee
The expanding rule are satisfied~\cite{Cahill}
\be
\label{rule}
\hat{T}(\alpha, s)=\frac{2}{t-s}\int\exp\{\frac{-2\vert \alpha-\beta\vert^2}{t-s}\}\hat{T}(\beta,t)\frac{d^2\beta}{\pi}.
\ee
In the so-called P-representation~\cite{Scully}, a density matrix $\hat{\rho}$ can be expressed in terms of the coherent states,
\begin{equation}
\hat{\rho}=\int P(\beta,\beta^*)\vert \beta\rangle\langle\beta\vert \frac{d^2\beta}{\pi},
\end{equation}
where $P(\beta,\beta^*)=\mathrm{Tr}[\hat{\rho}\delta(\beta^*-\hat{a}^{\dagger})\delta(\beta-\hat{a})]$ satisfies the normalization condition
\be
\label{normaliztion}
\int P(\beta,\beta^*)\frac{d^2\beta}{\pi}=1.
\ee

Next, we can consider the case where the ensemble of inputs consists of coherent states $\vert \alpha\rangle$ with $\alpha$ distributed by the Gaussian distribution
\be
p(\alpha)=\eta\exp\{-\eta\vert\alpha\vert^2\},
\ee
with $\eta^{-1}$ the inverse width. The corresponding target states is defined to be $\vert g\alpha\rangle$. Using Eq.~(\ref{rule}), the density operator
$\hat{\tau}$ can be expressed as
\be
\hat{\tau}:&=&\int p(\alpha) \vert \alpha\rangle\langle  \alpha\vert\frac{d^2\alpha}{\pi},\nonumber\\
&=&\hat{T}(0,-\frac{\eta+2}{\eta}),
\ee
and the Sqrt POVM associated with the input ensemble is
\be
\label{sqrt1}
\hat{S}(\alpha)=(\eta+1)\vert \sqrt{1+\eta}\alpha\rangle\langle\sqrt{1+\eta}\alpha\vert.
\ee
By jointing it with the definition of the target states, the process matrix can be obtained
\be
\lambda_{\varepsilon_{\mathrm{EB}}}:&=&\int\frac{d^2\alpha}{\pi}\vert \vert g\alpha\rangle\langle g\alpha\vert\rangle\rangle\langle\langle \hat{S}(\alpha)\vert,
\nonumber\\
&=&\int\frac{d^2\alpha}{\pi}\vert \vert \kappa\alpha\rangle\langle \kappa\alpha\vert\rangle\rangle \langle\langle \vert\alpha\rangle\langle \alpha\vert \vert,
\ee
where the coefficient $\kappa$ is
\be
\kappa=\frac{g}{\sqrt{1+\eta}}.
\ee
With the multiplication law in Eq.~(\ref{mutiplication}), one may verify that the process matrix is invariant under the unitary decomposition with $U=\hat{D}(\beta)$ and $V=\hat{D}(\kappa\beta)$, say
\be
\hat{\lambda}_{\varepsilon_{\mathrm{EB}}}=\hat{D}({\kappa\beta})\otimes\hat{D}^*({\kappa\beta})  \lambda_{\varepsilon_{\mathrm{EB}}} [\hat{D}((\beta)\otimes
 \hat{D}^*(\beta)]^{\dagger}.
 \ee
According to Eq.~(\ref{relation}), we have
\be
\label{result}
\vert\vert\varepsilon_{\mathrm{EB}}(\vert\beta\rangle\langle\beta\vert)\vert\vert_{\infty}=\vert\vert\varepsilon_{\mathrm{EB}}(\vert 0\rangle\langle 0 \vert)\vert\vert_{\infty}.
\ee

Now, take an arbitrary density matrix $\hat{\rho}$ as the input of the effective EB channel, and after evolution, the final state becomes
\be
\label{final}
\varepsilon_{\mathrm{EB}}(\hat{\rho})=\int P(\beta,\beta^*) \varepsilon_{\mathrm{EB}} (\vert \beta\rangle\langle\beta\vert) \frac{d^2\beta}{\pi},
\ee
For two density operators $\hat{\rho}_1$ and $\hat{\rho}_2$, it was shown in Ref.~\cite{Fuchs} that
\be
\vert\vert(\hat{\rho}_1+\hat{\rho}_2)\vert\vert_{\infty}\leq  \vert\vert \hat{\rho}_1\vert\vert_{\infty}+\vert\vert\hat{\rho}_2\vert\vert_{\infty}.
\ee
By jointing it with Eq.~(\ref{normaliztion}), Eq.~(\ref{result}),  and Eq.~(\ref{final}), the operator norm of $\varepsilon_{\mathrm{EB}}(\hat{\rho})$ can be obtained
\be
\vert\vert\varepsilon_{\mathrm{EB}}(\hat{\rho})\vert\vert_{\infty}&=& \vert\vert\int P(\beta,\beta^*) \varepsilon_{\mathrm{EB}} (\vert \beta\rangle\langle\beta\vert) \frac{d^2\beta}{\pi}\vert\vert_{\infty},      \nonumber\\
&\leq&\int P(\beta,\beta^*) \vert\vert\varepsilon_{\mathrm{EB}} (\vert \beta\rangle\langle\beta\vert)\vert\vert_{\infty} \frac{d^2\beta}{\pi},\nonumber\\
&=&\int P(\beta,\beta^*) \vert\vert\varepsilon_{\mathrm{EB}} (\vert 0\rangle\langle 0\vert)\vert\vert_{\infty} \frac{d^2\beta}{\pi},\nonumber\\
&=&\vert\vert\varepsilon_{\mathrm{EB}} (\vert 0\rangle\langle 0\vert)\vert\vert_{\infty}.
\ee
One can have the probabilistic CFT
\be
F^{\mathrm{prob}}_\mathrm{c}=\vert\vert\varepsilon_{\mathrm{EB}} (\vert 0\rangle\langle 0\vert)\vert\vert_{\infty}.
\ee
According to Eq.~(\ref{rule}), it can be found that $\varepsilon_{\mathrm{EB}} (\vert 0\rangle\langle 0\vert)$ is a thermal state
\be
\varepsilon_{\mathrm{EB}} (\vert 0\rangle\langle 0\vert)=\hat{T}(0,-(2\kappa^2+1)),
\ee
with the mean-number of quanta $\langle n\rangle=\kappa^2$.
Using Eq.~(\ref{thermal}) and Eq.~(\ref{mean}), one can come to
\be
F^{\mathrm{prob}}_\mathrm{c}=\frac{1+\eta}{1+\eta+g^2},
\ee
which was given by Chiribella and Xie. This probabilistic CFT is a tight-bound of the deterministic one, and for example, it can be attained with the Sqrt POVM defined in Eq.~(\ref{sqrt1}).

\section{discussions and conclusions}
\label{sec6}
The  target state is assumed to be the same as the input ones in Fhchs-Sasaki protocol~\cite{Fuchs}, but this is not our requirement in the present work. Besides, the probabilistic CFT is introduced in the reformulated Fuchs-Sasaki protocol, and it is showed that the CFTs can be defined in terms of the effective EB channel as in Ref.~\cite{Chir1}. However, instead of the Choi matrix used in
Ref.~\cite{Chir1}, the process matrix of the effective channel is used in our work for the derivation of CFTs. A series of examples are given to show the invariant property of the process matrix under the unitary decomposition, which has been discussed in Sec.~\ref{sec3} with an explicit form, and this property plays an important role in calculating CFTs.

The qubit states and coherent states are mainly focused in our work, and most of the examples have appeared in previous works. In recent works~\cite{Chir,Yang}, a series of probabilistic CFTs have been obtained for the cases where both the input and target states are neither the qubit states nor the coherent states. For such cases, how to decide the deterministic CFTs based on the reformulated Fhchs-Sasaki protocol, will be our future work.

Finally, let us end our work with a short conclusion. Following the idea of Chiribella and Xie, the protocol developed by Fuchs and Sasaki for  calculating CFTs, is reformulated in terms of the effective EB channel. The benchmark was determined by the process matrix of the effective EB channel. With the qubit states and coherent states as examples, it is shown that the reformulated protocol can be used as a convenient tool for CFTs.
 
\acknowledgements
This work was supported by the National Natural Science Foundation of China under Grants No.~11405136 and No.~11747311, and the Fundamental Research Funds for the Central Universities under Grant No.~2682016CX059.

\end{document}